\documentclass[prd,aps,showpacs,nofootinbib]{revtex4}
\usepackage{amssymb}
\usepackage{graphicx}
\usepackage{amsmath}

\newcommand{\be}{\begin{equation}}
\newcommand{\ee}{\end{equation}}
\newcommand{\bq}{\begin{eqnarray}}
\newcommand{\eq}{\end{eqnarray}}

\begin{document}

\title{Lorentz-symmetry Violation and Electrically
Charged Vortices in the Planar Regime}
\author{H. Belich Jr.$^{a,e}$,T. Costa-Soares$^{c,d,e}$, M.M. Ferreira Jr.$%
^{b,e}$, J.A. Helay\"{e}l-Neto$^{c,e},$M.T.D. Orlando$^{a,e}$}
\affiliation{$^{a}${\small {Universidade Federal do Esp\'{\i}rito Santo (UFES),
Departamento de F\'{\i}sica e Qu\'{\i}mica, Av. Fernando Ferrari, S/N
Goiabeiras, Vit\'{o}ria - ES, 29060-900 - Brasil}}}
\affiliation{$^{b}${\small {Universidade Federal do Maranh\~{a}o (UFMA), Departamento de F%
\'{\i}sica, Campus Universit\'{a}rio do Bacanga, S\~{a}o Luiz - MA,
65085-580 - Brasil}}}
\affiliation{{\small {~}}$^{c}${\small {CBPF - Centro Brasileiro de Pesquisas F\'{\i}%
sicas, Rua Xavier Sigaud, 150, CEP 22290-180, Rio de Janeiro, RJ, Brasil}}}
\affiliation{$^{d}${\small {Universidade Federal de Juiz de Fora (UFJF), Col\'{e}gio T%
\'{e}cnico Universit\'{a}rio, av. Bernardo Mascarenhas, 1283, Bairro F\'{a}%
brica - Juiz de Fora - MG, 36080-001 - Brasil}}}
\affiliation{$^{e}${\small {Grupo de F\'{\i}sica Te\'{o}rica Jos\'{e} Leite Lopes, C.P.
91933, CEP 25685-970, Petr\'{o}polis, RJ, Brasil}}}
\email{belich@cce.ufes.br, tcsoares@cbpf.br, manojr@ufma.br,
helayel@cbpf.br, orlando@cce.ufes.br}

\begin{abstract}
We start from a Lorentz non-invariant Abelian-Higgs model \ in 1+3
dimensions, and carry out its dimensional reduction to $D=1+2$. The planar
model resulting thereof is composed by a Maxwell-Chern-Simons-Proca gauge
sector, a massive scalar sector, and a mixing term (involving the fixed
background, $v^{\mu }$) that realizes Lorentz violation for the reduced
model. Vortex-type solutions of the planar model are investigated,
revealing charged vortex configurations that recover the usual
Nielsen-Olesen configuration in the asymptotic regime. The Aharonov-Casher
Effect in layered superconductors, that shows interference of neutral
particles with a magnetic moment moving around a line charge, is also
studied. Our charged vortex solutions exhibit a screened electric field that
induces the same phase shift as the one caused by the charged wire.
\end{abstract}

\maketitle

\section{ \ Introduction}

In recent years, some models with Lorentz and CPT breaking have been adopted
as a low-energy limit of an extension of the Standard Model, valid at the
Plank scale \cite{Kostelecky1},\cite{Colladay},\cite{Coleman}. An effective
action is obtained that incorporates CPT and Lorentz violation and keeps
unaffected the $SU(3)\times SU(2)\times U(1)$ gauge structure of the
underlying theory. Lorentz covariance is broken only in the particle frame,
remaining as a good symmetry at the point of view of the observer frame. A
well known way to introduce such a breaking in a QED, due to Carroll, Field
and Jackiw \cite{Jackiw}, consists in adopting a 4-dimensional version of
the Chern-Simons topological term, namely $\epsilon _{\mu \nu \kappa \lambda
}v^{\mu }A^{\nu }F^{\kappa \lambda }$, where $\epsilon _{\mu \nu \kappa
\lambda }$\ is the 4-dimensional Levi-Civita symbol and $v^{\mu }$\ is a
fixed four-vector acting as a background. This idea and consequences have
been extensively analyzed in refs. \cite{Adam}, \cite{Chung}.

Despite the intense activity proposing and analyzing the consequences of a
Lorentz-violating electrodynamics, experimental data and theoretical
considerations both indicate stringent limits on the parameters responsible
by such a breaking \cite{Coleman}, \cite{Particles} in a factual
(1+3)-dimensional electrodynamics. Such evidence raises the question about
the feasibility of observation of such effect in a low dimension system and
also in an environment distinct from the usual high-energy domain in which
this matter has been generally considered so far. It encourages the
investigation of the Lorentz-violation phenomenon in Condensed Matter planar
systems, for instance. Condensed Matter Systems (CMS), despite belonging to
the domain of low-energy physics, are \ known to exhibit some phenomena well
addressed by the mathematical tools of the high-energy physics (field
theory) domain, as it occurs in the case of the fractional Hall effect
(anyon excitations), kinks, vortices, and other topological defects. Such a
correspondence shows that there are some advantages in adopting a Field
Theory framework to address CMS's under different perspectives. It is well
known that CMS are sometimes endowed with spatial anisotropy which might
constitute a nice environment to study Lorentz-violation and to observe
correlated effects. Indeed, although Lorentz covariance is not defined in a
CMS, Galileo covariance holds as a genuine symmetry in such a system (a
least for the case of isotropic low-energy systems). Having in mind that a
CMS may be addressed as the low-energy limit of a relativistic model, there
follows a straightforward correspondence between the breakdown of Lorentz
and Galileo symmetries, in the sense that a CMS with\ violation of Galileo
symmetry may have as counterpart a relativistic system endowed with breaking
of Lorentz covariance.\thinspace\ Considering the validity of this
correspondence, it turns out that anisotropic CMS may be addressed as the
low-energy limit of a relativistic model in the presence of a spacelike
Lorentz-violating background.

\ The attainment of an attractive electron-electron $(e^{-}e^{-})$\
potential in the context of a model incorporating Lorentz-violation, for
instance, is a point that may set up a clear connection between such
theoretical models and condensed matter physics. Such issue has been already
taken into account both in (1+3) \cite{Altschul} and (1+2) dimensions \cite%
{Manojr4}. In two dimensions, it was adopted as starting point the planar
model corresponding to the dimensionally reduced version of the
Carroll-Field-Jackiw electrodynamics \cite{Manojr1}, \cite{Manojr2}. The
electron-electron interaction potential, obtained both for a purely timelike
and spacelike Lorentz-violating background \cite{Manojr4}, has revealed to
be attractive for some radial range, which leads to the real possibility of
attaining the formation of electron-electron bound states. Particularly, for
the case a purely spacelike background, the resulting interaction potential
is endowed with spacial anisotropy, once the Lorentz-violating two-vector
(v) sets up a fixed direction in space. Therefore, this study may be seen as
a first connection with planar superconductivity phenomena, including the
possibility of considering the anisotropy of the parameter of order as
related with the background.

In the context of\ a (1+3)-dimensional CMS, a fixed four-vector - $v^{\mu
}=( $v$_{0},$v), \ acting as a Lorentz-violating background,\ may be used to
describe anisotropy of a particular material, once it can select a
preferential plane in which phenomena such as planar superconductivity and
Quantum Hall Effect take place. In this way, some informations of the matter
environment may be associated with the geometric orientation of the fixed
four-vector. It is well known that superconducting ceramics are strong
anisotropic materials, in which the charge transport is essentially
two-dimensional. A suitable choice of the background spacial orientation can
be used to simulate anisotropy of such materials. Indeed, if one considers
the z-component of the four-vector orthogonal to the conducting plane, the x
and y spacial components will lay parallel to this plane. One has thus a
scenario in which the component v$_{z}$ defines the conduction plane,
whereas v$_{x}$\ and v$_{y}$\ are useful to simulate anisotropy on such a
plane. \ This is a general prescription for addressing four dimensional
systems in the presence of a Lorentz-violating background.

It should be still remarked that\ in Lorentz-violating theories endowed with
a fixed background, it is understood that such a background must be kept
invariant under Lorentz transformations in the particle frame \cite%
{Kostelecky1}, fact to which the breakdown of the covariance is
intrinsically related to. Therefore, we can effectively propose a gauge
theory suitable to such a requirement, that is, for which the fixed
background does not undergo Lorentz transformations (basically, rotations).
Generally, in CMS investigations one does not move the bulk, which may be
associated with an occasional background; only the fields are allowed to
change. Such a\ general description establishes a good compatibility between
CMS\ and our theoretical general conditions which define the Lorentz
violation.

Presently, the investigation of vortex configurations is of interest of both
Field Theory and Condensed Matter physics. This interplay was born since the
pioneering work of Nielsen \& Olesen \cite{Nielsen} that argued the
existence of (chargeless) vortices in the Abelian Higgs model (the
relativistic version of the Landau-Ginzburg theory, suitable for addressing
superconducting systems in a Field Theory approach). The presence of stable
(chargeless) vortex solutions in a non-Abelian Higgs context was soon
demonstrated as well \cite{Vortex1}. Some time later, the introduction of
the Chern-Simons term\ provided \ the attainment of charged vortex solutions
in the context of both Abelian \cite{Vortex2}\ and non-Abelian Higgs model 
\cite{Vortex3}, which are characterized by a linear relation involving the
charge and magnetic flux of the vortex.\ In the sequel, it was shown that
these vortex configurations satisfy a set of self-duality equations in the
pure anyonic limit \cite{Vortex4}, in which the dynamics of the gauge field
is ruled only by Chern-Simons term.

\ In the present work, we are particularly interested in studying vortex
solutions in the context of a Lorentz-violating Higgs Abelian model in (1+2)
dimensions. \ One possibility to simulate this kind of system is taking a
spatial component of $v^{\mu }$\ perpendicular to the plane in which the
supercurrent runs, according to the description given above. Another way
involves the definition of a planar theoretical model for which the third
spacial component is gotten rid of, although it provides a remanent in the
two-dimensional world. This is accomplished by means of the dimensional
reduction prescription in which the third spacial component is frozen. Such
a procedure will be suitably elucidated in Sec. II. The dimensional
reduction procedure is then implemented on the recently developed
Abelian-Higgs version of Carroll-Field-Jackiw electrodynamics \cite{Belich}.
The resulting (1+2)-dimensional model \cite{Manojr3} consists of a planar
electrodynamics composed of a Maxwell-Chern-Simons-Proca gauge field, two
scalar fields, and a mixing term (responsible for the violation of Lorentz
symmetry). The consistency of this model was also classically analyzed,
revealing that the causality, unitarity and stability of the modes are
assured for both purely timelike and spacelike backgrounds. This planar
model constitutes our theoretical framework for studying vortex-like
configurations in a Lorentz- violating Higgs-Abelian framework. It
corresponds to the relativistic analogue of the Landau-Ginzburg free energy,
which describes superconductivity \cite{Ryder} and vortex solutions in
superconductors \cite{Abrikosov}, here\ supplemented by the
Lorentz-violating mixing term. This model differ from the usual
Abelian-Higgs model analyzed by Paul \& Khare \cite{Vortex2} by the presence
of two terms containing the $\varphi -$\ scalar field (the remanent of the $%
A^{\left( 3\right) }$ component). Working with the resulting differential
equations, it is possible to show that the $\varphi $ field exhibits a
exponential decaying asymptotic behavior, limit in which our model recovers
the usual Nielsen-Olesen vortex configurations.

The organization of our work is as follows: Section\ II begins with a brief
description of fundamental mechanisms in low-temperature superconductivity\
in connection with the scenario in which the planar charged vortices appear.
Next, we perform\ the dimensional reduction of the
Maxwell-Chern-Simons-Higgs Electrodynamics. In Section III, we propose the
investigation of vortex-like solutions, discuss its electric charge and the
possibility of the appearance of a phase difference as a consequence of
their non-neutrality.\ In Section IV, we propose the discussion of the
interaction of particles in a superconducting medium with vortices and the
natural appearance of the Aharonov-Casher effect.\ \ Finally, in Sec. V, \
we present our Concluding Comments and Perspectives, where we also present a
possible connection of our theoretical model, vortices and Aharonov-Casher
effect in the so-called pseudo-gap regime of planar superconductors.

\section{The Dimensionally-Reduced Model}

One starts from the Maxwell Lagrangian in 1+3 dimensions supplemented with
the Carroll-Field-Jackiw term and a scalar sector endowed with spontaneous
symmetry breaking, as it appears in ref. \cite{Belich}: 
\begin{equation}
\mathcal{L}_{1+3}=\{-\frac{1}{4}F_{\hat{\mu}\hat{\nu}}F^{\hat{\mu}\hat{\nu}}-%
\frac{1}{4}\varepsilon ^{\hat{\mu}\hat{\nu}\hat{\kappa}\hat{\lambda}}v_{\hat{%
\mu}}A_{\hat{\nu}}F_{\hat{\kappa}\hat{\lambda}}+(D^{\hat{\mu}}\phi )^{\ast
}D_{\hat{\mu}}\phi -V(\phi ^{\ast }\phi )+A_{\hat{\nu}}J^{\hat{\nu}}\},
\label{action1}
\end{equation}%
where the $\hat{\mu}$ runs from $0$ to $3,$ $D_{\hat{\mu}}\phi =(\partial _{%
\hat{\mu}}+ieA_{\hat{\mu}})\phi $ is the covariant derivative, and $V(\phi
^{\ast }\phi )=m^{2}\phi ^{\ast }\phi +\lambda (\phi ^{\ast }\phi )^{2}$
represents the scalar potential responsible for the spontaneous symmetry
breaking ($m^{2}<0$ and $\lambda >0$)$.$ This model is gauge invariant but
does not preserve the Lorentz and CTP symmetries, once the fixed 4-vector, $%
v^{\hat{\mu}}$, does not undergo Lorentz transformation (it can be viewed as
a set of four scalars). \ It should be remarked that the spontaneous
symmetry breaking yields a Proca mass to the photon and induces the
transition to a non-trivial vacuum. Therefore, the light propagation in such
a material medium constitutes a short range interaction.\ This background
fluid, provided by the Higgs mechanism in a coherent way, does not suffer
scattering or phase decoherence by components of the crystalline lattice.
This scenario describes, actually, BSC superconductivity.

To study the planar counterpart of this model, one performs its dimensional
reduction to 1+2 dimensions, which consists effectively in adopting the
following ansatz over any 4-vector: (i) one keeps unaffected the temporal
and also the first two spatial components; (ii) one freezes the third
spacial dimension by splitting it from the body of the new 3-vector and
requiring that the new quantities $\left( \chi \right) $, defined in 1+2
dimensions, do not depend on the third spatial dimension:\ $\partial
_{_{3}}\chi \longrightarrow 0.$ It is also advisable to remark here that the
prescription adopted for the dimensional reduction ensures automatically
gauge invariance in (1+2) dimensions. Actually, the scalar, $s,$ associated
to $A^{\left( 3\right) }$ is a gauge-invariant scalar ($\partial _{3}\alpha
=0,$ $\alpha $ being the gauge parameter), where as $A^{\mu }($ $\mu =0,1,2)$
$\ $is subject to the usual gauge $U(1)$ transformation with parameter $%
\alpha .$ So, gauge invariance does match with our dimensional reduction
procedure. Other possible dimensional reduction schemes could be eventually
adopted, such as spontaneous compactification of the third spatial component 
\cite{Duff}. However, this particular reduction procedure introduces
non-zero massive modes which do not fit the phenomenology of excitations
appearing in CMS. It brings about the so-called towers of massive modes and
this is not pertinent with the physics of CMS. Actually, it seems that, if
one wishes to build field-theoretic models to describe low-dimensional CMS,
the procedure we choose to carry out here appears as the most suitable one.
A Legendre-type reduction could also be an interesting proposal to produce
models in lower dimensions \cite{Sohn}. It avoids the non-zero massive
models and it may give a new insight to the problem of the dependence on the
third space direction.

\ Now, applying\ our adopted prescription\ to the gauge 4-vector, $A^{\hat{%
\mu}},$ and the fixed external 4-vector, $v^{\hat{\mu}},$ one \ has: 
\begin{align}
A^{\hat{\nu}}& \longrightarrow (A^{\nu };\text{ }\varphi ) \\
v^{\hat{\mu}}& \longrightarrow (v^{\mu };\text{ }s),
\end{align}%
where: $A^{\left( 3\right) }=\varphi ,$ $v^{\left( 3\right) }=s$ and $\mu
=0,1,2$ ( note that all indices without hat have this range). According to
this process, there appear two scalars: the scalar field, $\varphi ,$ that
exhibits dynamics, and $s,$ a constant scalar (without dynamics). Carrying
out this prescription for eq. (\ref{action1}), one then obtains: \ \ \ \ \ \ 
\begin{align}
\mathcal{L}_{1+2}& =-\frac{1}{4}F_{\mu \nu }F^{\mu \nu }-\frac{s}{2}%
\varepsilon _{\mu \nu k}A^{\mu }\partial ^{\nu }A^{k}+\varphi \varepsilon
_{\mu \nu k}v^{\mu }\partial ^{\nu }A^{k}+\frac{1}{2}\partial _{\mu }\varphi
\partial ^{\mu }\varphi +(D^{\mu }\phi )^{\ast }D_{\mu }\phi -e^{2}\varphi
^{2}\phi ^{\ast }\phi  \notag \\
& -V(\phi ^{\ast }\phi )+A_{\nu }J^{v}-\varphi J,  \label{Lagrange2}
\end{align}%
The scalar field, $\varphi ,$ exhibits a typical Klein-Gordon massless
dynamics behavior and it also appears as the parameter that couples $v^{\mu
} $ to the gauge sector of the model by means of the Lorentz-violating
mixing term:\ $\varphi \varepsilon _{\mu \nu k}v^{\mu }\partial ^{\nu
}A^{k}. $ The presence of the Chern-Simons term in Lagrangian (\ref%
{Lagrange2}), here with mass $s,$ will amount also to the breakdown of the
parity and time-reversal symmetries, a phenomenon already observed in planar
superconducting systems which behave as a two-dimensional Fermi liquid \cite%
{Maeno}.

The mass dimensions of the fields and parameters contained in Lagrangian (%
\ref{Lagrange2}) are the following: $\left[ \varphi \right] =\left[ A^{\mu }%
\right] =[\phi ]=1/2,\left[ s\right] =\left[ v^{\mu }\right] =1,\left[ e%
\right] =1/2.$\ One thus notes that the dimensional reduction procedure
changes the dimension of the fields and coupling constants defined in the
(1+2) space-time in relation to their (1+3) counterparts. Indeed, while the
four-dimensional U(1) coupling constant, $e_{4}$, is dimensionless, the
tree-dimensional one exhibits a $\left[ \text{mass}\right] ^{1/2}$dimension,
a usual result for planar systems, ascribed to a possible effect remanent of
the third spacial dimension\footnote{%
In high-Tc superconductors, this effect may be associated with the coupling
between \ parallel conduction planes, taken into account in some theoretical
models.}. In this sense, such constant is taken as an effective coupling ($%
e_{3}=e_{4}/\sqrt{l_{\perp }})$\ by some authors \cite{Kogan}, with $%
l_{\perp }$\ standing for a characteristic length orthogonal to the plane. \
It is also instructive to discuss the possible role of the fields contained
in Lagrangian (\ref{Lagrange2}) in the context of a CMS. The first two terms
define the well-known Maxwell-Chern-Simons electrodynamics, which finds many
applications in connection with topological defects, anyonic excitations,
etc... The scalar field $\phi $ represents a Higgs sector responsible for
the appearance of stable vortex configurations, as already analyzed in the
context of both Abelian and non-Abelian scalar electrodynamics \cite{Vortex1}%
-\cite{Vortex4}. In the absence of currents and with a vanishing $\varphi $%
-field $\left( \varphi =0\right) $, one obtains exactly the same planar
Lagrangian which provides the existence of charged vortices \cite{Vortex2}
and self-dual configurations \cite{Vortex4} in the absence of the Maxwell
term. In this context, we should point out that the Lorentz-violating mixing
term constitutes the key difference of our planar model and models already
analyzed in literature. It is also of importance to discuss the physical
role played by the fixed background, $v^{\mu },$ in this theoretical
framework. It acts inducing both Lorentz violation and spatial anisotropy%
\footnote{\textbf{Such anisotropy reflects the directional dependence of the
solutions on the angle between this vector and the point where the solutions
are being considered. The electron-electron interaction potential evaluated
for the case of a purely spacelike background \cite{Manojr4} exhibits direct
dependence on such angle, confirming this role of the background.}} in the
plane (in the case it is spacelike).\ Therefore, the fixed background may be
seen as a physical element able to promote not only a shift at the energy
spectrum of the system, but also anisotropy and even changes in the physical
behavior of the systems\footnote{%
In ref. \cite{Manojr2}, it was verified that the Lorentz-violating
background was able to change the long distance behavior of the electric
field interaction in a Maxwell-Chern-Simons electrodynamics. Indeed, the
screened scalar potential $(A_{0})$ has been replaced by a logarithmic
interaction in the presence of the background.}.

The equations of motions obtained from (\ref{Lagrange2}) are: 
\begin{align}
\partial _{\nu }F^{\mu \nu }& =-s\text{ }\varepsilon ^{\mu \nu \rho
}\partial _{\nu }A_{\rho }+\varepsilon ^{\mu \nu \rho }v_{\nu }\partial
_{\rho }\varphi +ie(\phi \partial ^{\mu }\phi ^{\ast }-\phi ^{\ast }\partial
^{\mu }\phi )+2e^{2}\phi ^{\ast }\phi A^{\mu }+J^{\mu },  \label{a1} \\
\left( \square +M_{A}^{2}\right) \varphi & =\epsilon _{\mu \nu k}v^{\mu
}\partial ^{\nu }A^{k}+J,  \label{a2} \\
(D^{\mu })^{\ast }D_{\mu }\phi & =-e^{2}\varphi ^{2}\phi -m^{2}\phi
-2\lambda \left\vert \phi \right\vert ^{2}\phi ,  \label{a3}
\end{align}%
where we have\textit{\ }$M_{A}^{2}=2e^{2}\phi ^{\ast }\phi .$ Note that, by
the eq. (\ref{a3}) we have a increasing in the minimum of $V(\phi ^{\ast
}\phi )$. This implies that we obtain a planar Landau-Ginzburg theory with
critical field stronger that if we would start with a planar model.

We can extract from (\ref{a1}) the followings modified Maxwell equations%
\textit{\ }without current: 
\begin{align}
\partial _{t}\overrightarrow{E}+\tilde{\nabla}B\text{ \ }& =-s%
\overrightarrow{\tilde{E}}-\left( \overrightarrow{\text{\~{v}}}\partial
_{t}\varphi -\text{v}_{0}\overrightarrow{\tilde{\nabla}}\varphi \right)
-ie(\phi \overrightarrow{\nabla }\phi ^{\ast }-\phi ^{\ast }\overrightarrow{%
\nabla }\phi )-2e^{2}\left\vert \phi \right\vert ^{2}\vec{A},  \label{M1} \\
\overrightarrow{\nabla }.\overrightarrow{E}\text{ }+\text{ }\frac{s}{2}B%
\text{ }& =-\overrightarrow{\text{v}}\times \overrightarrow{\nabla }\varphi
-ie(\phi \partial _{t}\phi ^{\ast }-\phi ^{\ast }\partial _{t}\phi
)-2e^{2}\left\vert \phi \right\vert ^{2}A_{0},  \label{M2}
\end{align}%
where a tilde on the operators and fields stands for the dual in 2
space-dimensions, given as follows: $\tilde{\nabla}_{i}=\varepsilon
_{ij}\nabla _{j}$, and $\tilde{E}_{i}=\varepsilon _{ij}E_{j}$.

With these equations, we can establish a discussion of vortex-like
configuration in $\left( 1+2\right) $ dimensions. In the next section, we
shall obtain a charged vortex solution with\ $A_{0}$ generating a screened
electric field. We discuss the stability of the vortex and recover the
traditional Nielsen-Olesen solution in $\left( 1+2\right) $ dimensions.

\section{A discussion on vortex-like configurations}

Abrikosov \cite{Abrikosov}\ showed that the Ginzburg-Landau theory admits
vortex solutions. On the other hand, Nielsen and Olesen \cite{Nielsen}%
{\LARGE \ }discovered that the relativistic Abelian Higgs model also
presents vortex solutions while\ the Higgs field represents the
superconducting order parameter. \ These vortex solutions are important in
the Landau-Ginsburg model of superconductivity since the static energy
functional for the relativistic Abelian-Higgs model coincides with the
non-relativistic Landau-Ginzburg free energy in the description of type II
superconductors. \ It is well-known that cuprate superconductors exhibit an
enhanced effective mass anisotropy $(M_{x}\approx M_{y}\ll M_{z}=M_{A})$. In
other words, the charges are confined to planes,\ in which the vortices can
build up a lattice and\ amount to the destruction of the coherence of the
Cooper pairs and the planar superconducting current. Therefore, vortex
configurations in high-T$_{c}$\ superconductors should be addressed by a
planar model.  Recently, it was also discovered the presence of charged
fluxons in high-Tc superconductors \cite{Kobayashi}, which enforces the
quest for an anisotropic planar theory able for yielding charged vortices.\ 

To analyze the vortex-type solutions, we consider the charged scalar field
in $2$-dimensional space. Its asymptotic solution is proposed to be a circle 
$(S^{1})$ 
\begin{equation}
\phi=ae^{in\theta};~~~~~~(r\rightarrow\infty),  \label{38}
\end{equation}
where $r$ and $\theta$ are polar coordinates in the plane, $a$ is a constant
and $n$ is an integer.\textit{\ }In this limit we have\textit{\ }$%
M_{A}^{2}=2e^{2}a^{2}.$

We intend to study the asymptotic limit, and in this region the magnetic
field is vanishing. Then, the eq. (\ref{a2}), considering that no current
term is present, becomes:

\begin{equation*}
\left( \square +M_{A}^{2}\right) \varphi =0
\end{equation*}%
and we have a Klein-Gordon equation for the $A^{\left( 3\right) }$\
component. In the static regime, we have:

\begin{equation*}
\left( \nabla ^{2}-M_{A}^{2}\right) \varphi =0
\end{equation*}%
The solution is $\varphi \simeq \frac{1}{r}\exp (-M_{A}$ $r),$ and
asymptotically, we have $\varphi =0.$

On the other hand, the gauge field takes the following asymptotic behavior: 
\begin{equation}
\mathbf{A}=\frac{1}{e}\mathbf{\nabla }(n\theta );~~~~~~(r\rightarrow \infty
),  \label{39}
\end{equation}%
or, in terms of its components: 
\begin{equation}
A_{r}\rightarrow 0,~~~A_{\theta }\rightarrow -\frac{n}{er}%
;~~~~~~(r\rightarrow \infty ).  \label{40}
\end{equation}

The breaking of Lorentz covariance prevents us from setting $A_{\mu }$ as a
pure gauge at infinity, as usually done for the Nielsen-Olesen vortices.
This means that{\large \ }$A_{0}=A_{0}(r)$, as $r\rightarrow \infty $. In
this limit, we consider that $\varphi \rightarrow 0.$ The asymptotic
behavior of $A_{0}$ shall be fixed by the field equations, as shown in the
sequel. Returning to our problem, in this situation, the magnetic field
presents a cylindrical symmetry and 
\begin{equation}
\phi =\chi (r)e^{in\theta }.  \label{phi}
\end{equation}

Notice that, when the dimensional reduction is carried out, we obtain a new
potential: 
\begin{equation*}
\mathcal{V}(\phi ^{\ast }\phi )=e^{2}\varphi ^{2}\phi ^{\ast }\phi +V(\phi
^{\ast }\phi ),
\end{equation*}%
but the analysis of stability sets the point $\left( \varphi =0,\chi =\sqrt{-%
\frac{m}{2\lambda }}=a\right) $ as a new minimum.

To avoid singularity for $r\rightarrow0,$ and to keep an asymptotic
solution, we adopt the limits 
\begin{equation}
\lim_{r\rightarrow0}\chi(r)=0
\end{equation}
and 
\begin{equation}
\lim_{r\rightarrow\infty}\chi(r)=a.
\end{equation}

Notice that, asymptotically, the system goes to a specific non-trivial
configuration of the charged scalar field, which characterizes the
U(1)-symmetry breaking, but this breaking occurs in all directions on the
plane when $r\rightarrow \infty .$\ This configuration does not break
Lorentz or rotational symmetry; Lorentz (or even rotational symmetry)
breaking is explicitly realized by the three\textbf{-}vector $v^{\mu }$,
which creates a privileged direction in the space-time and thus breaks the
symmetry at the Lagrangian level.

To search for vortices, we need to analyze the modified Maxwell equations in
the static\textit{\ }regime to obtain the corresponding differential
equations, and then consider the asymptotic limit. Therefore, we can
eliminate the terms that have time dependence in our differential equations 
\cite{Ryder}. Taking as starting point eq. (\ref{a3}), 
\begin{equation}
\left( \partial _{i}-ieA_{i}\right) ^{\ast }\left( \partial
_{i}-ieA_{i}\right) \phi -m^{2}\phi -2\lambda ^{2}\left\vert \phi
\right\vert ^{2}\phi =0,
\end{equation}%
adopting the field parametrization (\ref{phi}), and summing over the
components, it becomes 
\begin{equation}
\frac{1}{r}\frac{d}{dr}\left( r\frac{d\chi }{dr}\right) -\left[ \left( \frac{%
n}{r}-eA\right) ^{2}+(e^{2}\varphi ^{2}+m^{2})+2\lambda \chi
^{2}+e^{2}A_{0}^{2}\right] \chi =0,  \label{ce}
\end{equation}%
while the modified Maxwell equations (\ref{M2}) take on the form: 
\begin{equation}
{\nabla }^{2}A_{0}-\overrightarrow{\text{v}}\times \overrightarrow{\nabla }%
\varphi -\frac{s}{2}B\text{ }-2e^{2}\chi ^{2}A_{0}=0.  \label{MM2}
\end{equation}%
Also, in the asymptotic region, this equation takes is read as:

\begin{equation}
{\nabla }^{2}A_{0}-M_{A}^{2}A_{0}=0,
\end{equation}%
whose simple solution is $A_{0}\simeq \frac{1}{r}\exp (-M_{A}$ $r).$
Considering now the $\theta $-component of eq. (\ref{a1}), \ it is written
as follows: 
\begin{equation}
\frac{d}{dr}\left( \frac{1}{r}\frac{d}{dr}\left( rA\right) \right) -2e\chi
^{2}\left( \frac{n}{r}+eA\right) -s\frac{dA_{0}}{dr}-\text{v}^{0}\frac{d}{dr}%
\varphi =0,  \label{m2}
\end{equation}%
which asymptotically is reduced to the form below: 
\begin{equation}
\frac{d}{dr}\left( \frac{1}{r}\frac{d}{dr}\left( rA\right) \right)
-2ea^{2}\left( \frac{n}{r}+eA\right) =0.
\end{equation}%
We then recover the Nielsen-Olensen solution$:$ 
\begin{equation}
A(r)=-\frac{n}{er}-\frac{c}{e}K_{1}\left( \sqrt{2}a\left\vert e\right\vert
r\right) ,
\end{equation}%
where $c$ is a constant, and $K_{1}$ is the modified Bessel function.

We notice that, asymptotically, the complex scalar field $\phi =\chi
(r)e^{in\theta }$ \ goes to a \ non-trivial vacuum and becomes $\phi
=ae^{in\theta }$ . Then, the topology of the vacuum manifold is $S^{1}$. We
also remark that the asymptotic behavior of the fields $\varphi $ and $A_{0}$%
\ is dominated by a Yukawa-like solution, whose corresponding mass is given
by $M_{A}^{2}=2e^{2}a^{2}.$\ We can trace the screening of these fields back
to the spontaneous symmetry breaking mechanism.\textit{\ }

Following the results of ref \cite{Manojr3}, which states the
stability, causality and unitarity of the reduced model here adopted, we
conclude that the stability of the vortex configuration is already assured.\
Furthermore,\ we can note that, whenever $sB\neq 0$, $A_{0}$ must
necessarily be non-trivial, and an electric field appears along with the
magnetic flux. If this is the situation, in the asymptotic region, $A_{0}$
falls off exponentially.

The appearance of an electrostatic field attached to the magnetic vortex,
whenever $sB\neq 0$, is not surprising. Its origin may be traced back to the
Lorentz-breaking term: indeed, being a Chern-Simons-like term, the
electrostatic problem induces a magnetic field and the magnetostatic regime
demands an electric field too. So, a non-vanishing $A_{0}$, and therefore a
non-trivial $\mathbf{E}$ is a response to the Chern-Simons Lorentz-breaking
term. The presence of a non-vanishing electrostatic potential points to a
charged vortex. In the next section, we shall continue exploiting
this feature of vortex configurations in a Lorentz breaking theory.

\section{On the Aharonov-Casher Effect}

A question which becomes relevant in Quantum Mechanics is to understand how
to attribute to a phase factor in the wave function effects that classically
are trivial due to the vanishing of the force. The Aharonov-Bohm (AB) effect 
\cite{ABohm} is a remarkable example of how we can provide with a real
meaning a quantity that could, apparently, appear as simple a mathematical
artifact. In this effect, the electromagnetic vector potential induces a
change in the wave function associated to an electron moving in a region
where there is no magnetic field. In the work by Aharonov and Casher \cite%
{Aharanov}, we have a dual aspect of this effect. They predicted a phase
shift in the wave function of a neutral particle with magnetic dipole moment 
$\mathbf{\mu },$ due to the action of an external electrostatic field:

\begin{equation}
\Delta \Phi _{AC}=\frac{1}{\hbar c^{2}}\oint \mathbf{\vec{\mu}}\times 
\mathbf{\vec{E}.}d\mathbf{r,}  \label{fase}
\end{equation}%
with $E$ being the electric field applied to magnetic dipole due to a
charged wire (for the sake of calculating the phase, we have restored the
constants $c$ and $\hbar $, up to now taken both equal to 1 ). In the last
section of ref. \cite{Aharanov}, it is shown the term $\vec{V}\cdot \vec{E}%
\times \vec{\mu}$\ that generates no force and is responsible for generating
an A-B phase. In such a work, taking a Maxwell-Higgs model with spontaneous
symmetry breaking in $(1+2)$ dimensions, the vortex solution is presented,
but the charged wire is put in by hand. Our model (\ref{Lagrange2}), by
virtue of the Lorentz-breaking Chern-Simons term, has a richer structure for
the vortices, and an electrostatic potential $A^{0}\neq 0$ is a natural
solution from the equations of motion. Then, the effect of the charged wire
in the work of the ref. \cite{Aharanov} is, in our proposal, replaced by the
non-neutral vortex that is naturally formed as a response to the vector
background that breaks Lorentz symmetry by means of a Chern-Simons type
action term.

The vortex charge is viewed as a source and the electric field generated
seems to be screened by the environment $\left( A_{0}\simeq \frac{1}{r}\exp
(-M_{A}r)\right) $. Charged vortices in high-T$_{c}$ superconductors are
proposed in \cite{Khom} as consequence of the variation of the chemical
potential in the phase transition. The superconductor fluid screens all the moment
of the electric field, but does not screens the topological effect (\ref%
{fase}). Our purpose is to establish a phase shift from the interaction
between two vortices. Lets suppose a planar circular superconductor with two
vortices: one pinned at the center and another circulating around. To obtain
the corresponding effect in this case, we can identify $\mathbf{\vec{\mu}}%
\simeq \Phi _{0}\hat{z};$ in so doing, the analogous expression for the
phase of the circulating vortex is given as below:

\begin{equation}
\vec{V}\cdot \vec{E}\times \mathbf{\Phi _{0}\hat{z},}  \label{Phi}
\end{equation}%
where $\vec{V}$ is the vortex velocity, $\vec{E}$ is the electric field, $%
\hat{z}$ is the unit vector perpendicular to the plane, and $\Phi _{0}=\frac{%
hc}{q}$ is called a flux quantum. As it is well known, a high-T$_{c}$
superconductor may support quantized localized magnetic fluxons with a flux $%
\Phi _{0}$ in a fluid environment with unit charge $q=2e$. The idea of a
circular geometry for superconductors was first explored in ref. \cite%
{Reznik}. The proposal of introducing a charge potential vector that yields
a phase shift (in the same sense that a normal potential vector creates the
A-B phase) was suggested by van Wess in the work of ref. \cite{van}. This
charge potential vector can also produce a persistent voltage in a ring \
made up of arrays of Josephson junctions. The dynamics of a single vortex
present in a ring-shaped (Corbino geometry) two-dimensional array of
low-capacity Josephson junction is analyzed. The vortex is seen as a
macroscopic quantum particle, whose energy levels $E_{n}\left( Q_{0}\right) $
are periodic functions of the externally induced gauge charge $Q_{0}$ which
is enclosed by the vortex, with a period $2e.$ The quantization of the
voltage generated by ballistic vortices with a mass $m_{v}$ in a
two-dimensional superconductor ring is assessed in \cite{Orlando}. The
experimental observation of this quantum interference is verified in \cite%
{Elion} and it enforces the relevance of the study of charged-vortex
interaction. In our case, the configuration of the gauge field which
describes the vortex is steady. A possible connection amongst charged
vortex, A-C effect and planar superconductor will be pointed out in our next
section. 

\section{Concluding Comments and Perspectives}

In this work, we have studied a planar Abelian gauge models attained as a
by-product of a 4-dimensional theory under a dimensional reduction
process.  This procedure has provided reasonable results under both the
theoretical and phenomenological points of view, with the advantage that
some specific 3-dimensional aspects may be understood as a manifestation of
mechanisms or characteristics of the 4-dimensional system. In this sense,
for instance, the topological mass parameter ($s$) of the planar system is
nothing but the\ axial component of the 4-dimensional vector that breaks
Lorentz and CPT symmetries. In other words: topologically massive planar
gauge theories may be justified as an inheritance from the
Carroll-Field-Jackiw Lorentz-violating gauge theory in (1+3)-D. 

An interesting feature considered here is the appearance of electrically
charged magnetic vortices in the set-up of a planar gauge theory for which
the mass parameter $s$ is non-vanishing. Indeed, in the framework of this
model there appear charges vortex configurations that recover the
Nielsen-Olesen structure in the asymptotic limit. Once vortex configurations
are associated with a intrinsic electric field, it settles down the full
scenario for the study of Aharonov-Casher effect. It was then shown that an
A-Casher phase is induced whenever a spin-1/2 neutral particle move around
the vortex. The point is that our vortex configurations need not be
supplemented by a thin charged wire in order to incorporate the effect of a
charge. The screening\ of the field configurations and the net charge of the
vortex come naturally out as an inheritance of Maxwell-Chern-Simons sector
generated from the Carroll-Field-Jackiw term. 
As an example of possible properties of charged vortices in high-Tc
superconductors, we refer to the $Hg_{0.82}Re_{0.18}Ba_{2}Ca_{2}Cu_{3}O_{8+%
\delta }$. We are interested in investigating the properties of the planar
charged vortices before the phase transition to superconductivity in ceramic
superconductors: the so-called pseudo gap regime. The phase transition
diagram in high Tc-superconductors is quite complex \cite{Mello}. First of
all, we can say that there occurs the formation of puddles's
superfluid below the $T^{\ast }$\ (temperature of pseudo-gap transition).
When the temperature decreases, the frontiers of the puddles grow and
percolate. At this point, it settles down the formation \ of the
superconducting planes, and we can treat the system in the regime of a
macroscopic planar physics. The 2D character of the fluctuation spectrum in
this temperature region, which extends down to (T-Tc)/Tc of the order of
0.01, was justified by the strong planar anisotropy that characterizes the
Hg,Re-1223 high-Tc superconductors \cite{Roa}.

In this environment, below $T^{\ast }$ and down to $T_{c},$ the gap's
screening is not effective because we do not have a stabilized percolative
superfluid. The expression for the electric field generated by the charged
vortex is:

\begin{equation}
\vec{E}=\left( \frac{1}{r^{2}}+\frac{M_{A}}{r}\right) \exp (-M_{A}r)\hat{r},
\end{equation}%
\ and we have the Aharonov-Casher term (\ref{Phi}) promoting interaction
among vortices. If we take consider electrons moving in the region close to
the core of the vortex, which amounts to taking into account the
approximation $\exp (-M_{A}r)\simeq 1-M_{A}r,$\ we can compute the stable
phase factor (\ref{fase}): $\mathbf{\Delta \Phi }_{AC}\mathbf{=(}\Phi
_{0}/\hbar c^{2})\frac{\mathbf{2\pi }}{d},$ where $d$ stands for the
distance of the electron to the center of the vortex. Another interesting
scenario for the application appears when the superconducting regime is
established. The high-T$_{c}$ superconductors are built by planes containing
Cu and O, which are sandwiched between two planes: one containing Ba and O
and another containing Hg and O. The outers planes (Ba-O and Hg-O) present
isolant properties and the Cu-O planes are conducting ones \cite{Jorg}.
Linear combinations of Cu and O atomic orbitals, called "hybrid orbitals",
are used to form covalent bindings that hold those atoms together. For these
high-T$_{c}$ superconductors, the conducting planes are formed by $3d_{%
\mathbf{(x}^{2}\mathbf{-y}^{2}\mathbf{)}}-2p$ hybrid orbitals of Cu and O 
\cite{Ball}. The c-axis presents a different behavior, which can be compared
with semiconducting materials (the resistance is not linear as a function of
the temperature). According to Allen \textit{et al.} \cite{Allen}, the
fundamental physics of the oxyde superconductors is contained in the Hubbard
Hamiltonian on a two-dimensional square lattice for small numbers of holes.
In this kind of model, the superconducting planes interact by charge
tunneling through the z axis, which can be interpreted as a strong
anisotropy or low z-axis charge mobility. In this situation, the z-axis
charge can be modeled as a static charge, and the plane charge as a moving
charge. If we consider the z-direction formed by hybridized 3d$_{\mathbf{(z}%
^{2}\mathbf{)}}$-2p bound as a charged wire with linear densities, $\lambda $%
\ , the vortices can present \ a stable phase upon the circulation around
this wire given by (\ref{Phi}): $\Delta \Phi _{AC}=(\Phi _{0}/\hbar
c^{2})4\pi \lambda .$With these results, we have presented possible effects
that charged vortices could suffer in planar superconductivity, when it the
phase transition region T* down to Tc was considered.

In a recent work \cite{Thales}, it has been shown that a non-minimal
coupling of fermions to a Lorentz-violating background induces an
Aharonov-Casher phase. It is known that spinless neutral particles may
develop\ a magnetic dipole moment whenever non-minimally coupled to an
electromagnetic field. If this is the case, it would be worthwhile to
explicitly derived the Aharonov-Casher phase of a spinless neutral particle
placed in the electric field of a charged wire. We argue that the Lorentz
breaking in 4 D may induce the magnetic moment for spinless particles and
then, the reduced 3-dimensional system may indicate how the scalar particle
may pick up the Aharonov-Casher phase by means of its induced magnetic
moment. This issue is under investigation.

\section{ Acknowledgments}

One of the authors (H. Belich) expresses his gratitude to the High Energy
Section of The Abdus Salam ICTP for the kind hospitality during the period
this work was done. CNPq is also acknowledged for the invaluable financial
support.


\begin{thebibliography}{99}
\bibitem{Kostelecky1} V. A. Kostelecky and S. Samuel, Phys. Rev. \textbf{D} 
\textbf{39}, 683 (1989); V. A. Kostelecky and R. Potting, \emph{\ \ }Nucl.
Phys. \textbf{B359}, 545 (1991); ibid, Phys. Lett.\textbf{\ B381} , 89
(1996); V. A. Kostelecky and R. Potting, Phys. Rev. \textbf{D51} , 3923
(1995);

\bibitem{Colladay} D. Colladay and V. A. Kosteleck\'{y}, Phys. Rev. \textit{{%
\ \textbf{D}}} \textbf{55},6760 (1997); D. Colladay and V. A. Kosteleck\'{y}%
, Phys. Rev.\textit{{\ \textbf{D }}}\textbf{58}, 116002 (1998); S.R. Coleman
and S.L. Glashow, Phys. Rev. \textbf{D 59}, 116008 (1999).

\bibitem{Coleman} S.R. Coleman and S.L. Glashow, Phys. Rev. \textbf{D 59},
116008 (1999); V. A. Kosteleck\'{y} and M. Mewes, Phys. Rev. Lett. \textbf{\
87 }, 251304 (2001); Phys. Rev. \textbf{D 66}, 056005 (2002).

\bibitem{Jackiw} S. M. Carroll, G. B. Field and R. Jackiw, Phys. Rev. 
\textbf{D 41}, 4, 1231 (1990).

\bibitem{Adam} C. Adam and F. R. Klinkhamer, Nucl. Phys.\textit{{\ \textbf{B}%
}} \textbf{607}, 247 (2001); C. Adam and F.R. Klinkhamer, \emph{\ }Phys.
Lett. \textbf{B }513, 245 (2001) ; V.A. Kostelecky and R. Lehnert, Phys.
Rev. \textbf{D 63,} 065008 (2001) ; A.A. Andrianov, P. Giacconi and R.
Soldati, JHEP 0202, 030 (2002); A.A. Andrianov, R. Soldati and L. Sorbo,
Phys. Rev\textit{.} \textbf{D 59}, 025002 (1999).

\bibitem{Chung} R. Jackiw and V. A. Kosteleck\'{y}, Phys. Rev. Lett. \textbf{%
82}, 3572 (1999); J. M. Chung and B. K. Chung Phys. Rev\textit{. }\textbf{D} 
\textbf{63}, 105015 (2001); J.M. Chung, Phys.Rev. \textbf{D 60}, 127901
(1999); M. Perez-Victoria, Phys. Rev. Lett. \textbf{83} , 2518 (1999); G.
Bonneau,\emph{\ }Nucl.Phys. \textbf{B 593, }398 (2001); M. Perez-Victoria,
JHEP 0104, 032 (2001); A. P. Ba\^{e}ta Scarpelli \textit{et al}., \textit{%
Phys. Rev.} \textbf{D 64}, 046013 (2001).

\bibitem{Particles} E. O. Iltan, \textit{Mod. Phys. Lett}. \textbf{A19}, 327
(2004); -ibid, \textit{Mod. Phys. Lett}. \textbf{A19}, 2215 (2004).

\bibitem{Belich} A. P. Ba\^{e}ta Scarpelli, H. Belich, J. L. Boldo and J. A.
Helay\"{e}l-Neto, Phys. Rev.\emph{\ }\textbf{D} \textbf{67}, 085021 (2003).

\bibitem{Altschul} B. Altschul, Phys. Rev. \textbf{D} 70, 056005 (2004).

\bibitem{Manojr4} M. M. Ferreira Jr, Phys. Rev. \textbf{D} 70, 045013
(2004); -ibid, Phys. Rev. \textbf{D} 71, 045003 (2005).

\bibitem{Manojr1} H. Belich, M.M. Ferreira Jr., J. A. Helay\"{e}l-Neto and
M.T.D. Orlando, Phys. Rev.\emph{\ }\textbf{D} \textbf{67}, 125011 (2003);
-ibid, Phys. Rev. \textbf{D} \textbf{69}, 109903 (E) (2004).

\bibitem{Manojr2} H. Belich, M.M. Ferreira Jr., J. A. Helay\"{e}l-Neto and
M.T.D. Orlando, Phys. Rev.\emph{\ }\textbf{D} \textbf{68}, 025005 (2003).

\bibitem{Manojr3} H. Belich, M.M. Ferreira Jr., and J. A. Helay\"{e}l-Neto,
Eur. Phys. J. \textbf{C 38}, 511 (2005).

\bibitem{Nielsen} H. B. Nielsen and P. Olesen, Nucl. Phys. \textbf{B 61}, 45
(1973).

\bibitem{Vortex1} H. J. de Vega, Phys. Rev. \textbf{D 18}, 2932 (1978), P.
Hasenfratz, Phys. Lett. \textbf{B 85}, 338 (1979); A. S. Schwartz and Y. S.
Tyupkin, Phys. Lett. \textbf{B 90}, 135 (1980).

\bibitem{Vortex2} S. K. Paul and A. Khare, Phys. Lett. \textbf{B 174}, 420
(1986); Phys. Lett. \textbf{B 182}, 414 (E) (1986)); -ibid, Phys. Lett. 
\textbf{B 191}, 389 (1986).

\bibitem{Vortex3} H. J. de Vega and F. A. Schaposnik, Phys. Rev. Lett. 
\textbf{56}, 2564 (1986); C. N. Khumar abd A. Khare, Phys. Lett. \textbf{B
178}, 395 (1986).

\bibitem{Vortex4} J. Hong, Y. Kim, and P. Y. Pac, Phys. Rev. Lett. \textbf{64%
}, 2230 (1990); R. Jackiw and E. J. Weinberg, Phys. Rev. Lett. \textbf{64},
2234 (1990).

\bibitem{Ryder} "\textit{Quantum Field Theory}", L. H. Ryder, Cambridge
University Press, 1985.

\bibitem{Abrikosov} A.A. Abrikosov, Zh. Eksp. Teor. Fiz. \textbf{32}, 956
(1957) [Sov. JETP \textbf{5}, 1174 (1957)].

\bibitem{Duff} M.J. Duff, "\textit{Modern Kaluza-Klein Theories}", in
"Supersymmetry and Supergravity '84", Proc. of the Trieste Spring School,
ed. by B. de Wit, P. Fayet and P. van Nieuwenhuizen, World Scientific,
Singapore, 1984.

\bibitem{Sohn} M. F. Sohnius, K. S. Stelle and P.C. West, Nucl. Phys. 
\textbf{B }173 (1980) 127.

\bibitem{Maeno} Y. Maeno \textit{et al.}, Nature \textbf{372}, 532 (1994);
T.M. Rice and M. Sigrist, J. Phys.: Condens. Matter \textbf{7}, L643 (1995);
A. P. Mackenzie \textit{et al.}, Phys. Rev. Lett. \textbf{76}, 3786 (1996);
Y. Maeno, Physica \textbf{C} 282-287, (1997).

\bibitem{Kogan} Ya.I. Kogan, JETP Lett. \textbf{49}, 225 (1989); S.
Randjbar-Daemi \textit{et al.}, Nucl. Phys. \textbf{B 340}, 403 (1990).

\bibitem{Kobayashi} "\textit{Vortex Electronics and Squids}" (Topics in
Applied Physics) by Takeshi Kobayashi, H. Hayakawa, M. Tonouchi, T.
Kobayashi, Hisao Hayakawa, Masayoshi Tonouchi, Springer-Verlag, (February 1,
2004); D. I. Khomskii and A. Freimuth, Phys. Rev. Lett. \textbf{75}, 1384
(1995); G. Blatter \textit{et al}., Phys. Rev. Lett. \textbf{77}, 566
(1996); J. Kol\'{a}cek and P. Lipavsk\'{y}, Phys. Rev. Lett. \textbf{86},
312 (2001); Y. Chen, Z.D. Wang, J.-Win Zhu, and C. S. Ting, Phys. Rev. Lett. 
\textbf{89}, 217001 (2002).

\bibitem{ABohm} Y. Aharanov and D. Bohm, Phys. Rev. \textbf{B} \textbf{3, }%
485 (1959)\textbf{.}

\bibitem{Aharanov} Y. Aharanov and A. Casher, Phys. Rev. Lett. \textbf{53},
319 (1984).

\bibitem{Khom} D. I. Khomskii and A. Freimuth, Phys. Rev. Lett. \textbf{75},
1384 (1995).

\bibitem{Reznik} B. Reznik and Y. Aharonov, Phys. Rev.\emph{\ }\textbf{D} 
\textbf{40, }4178 (1989).

\bibitem{van} B. J. van Wess,\emph{\ }Phys. Rev. Lett. \textbf{65}, 25,
(1989); B. J. van Wess, Phys. Rev. Lett. \textbf{65}, 255 (1990).

\bibitem{Orlando} T. P. Orlando and K. A. Delin, Phys. Rev.\emph{\ }\textbf{%
\ B } \textbf{43, } 8717 (1991).

\bibitem{Elion} W. J. Elion, \ J. J. Wachters, L. L. Sohn, and J. E. Mooij,
Phys. Rev. Lett. \textbf{71}, 2311 (1993).

\bibitem{Mello} E. V. L. de Mello et all, Phys. Rev. B. \textbf{66},
092504-1, (2002).

\bibitem{Roa} J Roa-Rojas et. all, Sup. Sci. Tech. \textbf{14}, 898, (2001).

\bibitem{Jorg} J. D. Jorgensen et al, Phys. Rev. B \textbf{36}, 3608, (1987).

\bibitem{Ball} C. J. Ballhausen, "\textit{Introduction to Ligant Field",
Theory} MacGraw-Hill, New York, 1962.

\bibitem{Allen} P. B. Allen, "\textit{High-Temperature Superconductivity",}
Chap. 9, Springer-Verlag, Berlin, 1990.

\bibitem{Thales} H. Belich, T. Costa-Soares, M.M. Ferreira Jr., and J. A.
Helay\"{e}l-Neto, Eur. Phys. J. \textbf{C 41}, 421 (2005).
\end{thebibliography}
\end{document}